\documentclass[aps,prd,twocolumn,groupedaddress,eqsecnum,showpacs,nofootinbib]{revtex4}
\usepackage{graphicx}
\usepackage{bm}
\usepackage{dcolumn}
\usepackage{hyperref}
\usepackage{breakurl}

\usepackage{amsmath}
\usepackage{amssymb}

\numberwithin{equation}{section}
\usepackage{subfigure}

\def \lleq {\lower0.9ex\hbox{ $\buildrel < \over \sim$} ~}
\def \ggeq {\lower0.9ex\hbox{ $\buildrel > \over \sim$} ~}

\def \beq  {\begin{equation}}
\def \eeq  {\end{equation}}
\def \ber  {\begin{eqnarray}}
\def \eer  {\end{eqnarray}}
\def\bq{\begin{equation}}
\def\nq{\end{equation}}
\def\bqr{\begin{eqnarray}}
\def\nqr{\end{eqnarray}}

\begin{document}

\newcommand{\newc}{\newcommand}
\newc{\be}{\begin{equation}}
\newc{\ee}{\end{equation}}
\newc{\ba}{\begin{eqnarray}}
\newc{\ea}{\end{eqnarray}}
\newc{\bea}{\begin{eqnarray*}}
\newc{\eea}{\end{eqnarray*}}
\newc{\D}{\partial}
\newc{\daa}{\left(\frac{\Delta \alpha}{\alpha}\right)}
\newc{\ie}{{\it i.e.} }
\newc{\eg}{{\it e.g.} }
\newc{\etc}{{\it etc.} }
\newc{\etal}{{\it et al.}}
\newc{\lcdm}{$\Lambda$CDM}
\newc{\dmu}{\left(\frac{\Delta \mu (z)}{\bar \mu (z)}\right)}
\newcommand{\nn}{\nonumber}
\newc{\ra}{\rightarrow}
\newc{\lra}{\leftrightarrow}
\newc{\lsim}{\buildrel{<}\over{\sim}}
\newc{\gsim}{\buildrel{>}\over{\sim}}
\newcommand{\dslash}{D\!\!\!\!/}
\newcommand{\bx}{{\bf x}}
\newcommand{\bn}{{\bf n}}
\newcommand{\bk}{{\bf k}}
\newcommand{\dd}{{\rm d}}
\def\ga{\mathrel{\raise.3ex\hbox{$>$\kern-.75em\lower1ex\hbox{$\sim$}}}}
\def\la{\mathrel{\raise.3ex\hbox{$<$\kern-.75em\lower1ex\hbox{$\sim$}}}}

\title{CMB Maximum Temperature Asymmetry Axis: Alignment with Other Cosmic Asymmetries}
\author{Antonio Mariano}\email{antonio.mariano@le.infn.it}
\affiliation{Department of Mathematics and Physics,
University of Salento \& INFN, Via Arnesano, 73100 Lecce, Italy}

\author{Leandros Perivolaropoulos}
\email{leandros@uoi.gr}
\affiliation{Department of Physics, University of Ionnina, Greece}

\date{\today}

\begin{abstract}
We use a global pixel based estimator to identify the axis of the
residual Maximum Temperature Asymmetry (MTA) (after the dipole
subtraction) of the WMAP 7 year Internal Linear Combination (ILC) CMB
temperature sky map. The estimator is based on
considering the temperature differences between opposite pixels in the
sky at various angular resolutions ($4^\circ-15^\circ$ and selecting
the axis that maximizes this difference. We consider three large scale
Healpix resolutions ($N_{side}=16$ ($3.7^\circ$), $N_{side}=8$
($7.3^\circ$) and $N_{side}=4$ ($14.7^\circ$)). We compare the
direction and magnitude of this asymmetry with three other cosmic
asymmetry axes ($\alpha$ dipole, Dark Energy Dipole and Dark Flow) and
find that the four asymmetry axes are abnormally close to each
other. We compare the observed MTA axis with the corresponding MTA
axes of $10^4$ gaussian isotropic simulated ILC maps (based on
\lcdm). The fraction of simulated ILC maps that reproduces the
observed magnitude of the MTA asymmetry and alignment with the
observed $\alpha$ dipole is in the range of $0.1\%-0.5\%$ (depending on the resolution
chosen for the CMB map). The corresponding magnitude+alignment probabilities
with the other two asymmetry axes (Dark Energy Dipole and Dark Flow)
are at the level of about $1\%$. We propose Extended Topological
Quintessence as a physical model qualitatively consistent with this
coincidence of directions.
\end{abstract}

\maketitle
\section{Introduction}
There is observational evidence coming mainly from the isotropy of the
Cosmic Microwave Background (CMB) that the Universe is isotropic on
Hubble scales. Any anisotropy on these scales is bound to be smaller
than about 1 part in $10^3$. This constraint combined with the
Copernican principle (supported by kSZ data \cite{Zhang:2010fa}) leads
to strong support of the cosmological principle: the Universe is
homogeneous and isotropic on Hubble scales.

The violation of the cosmological principle is expected to occur at a
small level even on Hubble scales due to small statistical fluctuations
of the cosmic energy density (matter, radiation, dark energy). Precise
cosmological observations are in principle able to detect this
deviation from the cosmological principle on Hubble scales and compare
the expected magnitude with the one anticipated based on the standard
isotropic cosmological model.

The lowest order deviation from isotropy, which is also easiest to
detect, is the anisotropy that distinguishes a preferred cosmological
axis. Such an axis is usually reasonably described by a dipole
deviation from isotropy. An exception consists of the CMB temperature
perturbations where the dipole term is dominated by our motion with
respect to the CMB frame and therefore it has been removed completely
from the CMB maps. This removal has also swept away any subdominant
cosmological contribution to the dipole. However any axial
cosmological anisotropy that is not perfectly described by a dipole
could have left a trace after the removal of the dipole. The detection
of this trace may be possible by using specially designed statistical
tests.

Early hints for deviations from isotropy on Hubble scales have been
accumulating during the last decade. Some of these hints may be
summarized as follows\cite{Antoniou:2010gw,Perivolaropoulos:2008ud}:
\begin{enumerate}
\item
{\bf Large Scale Velocity Flows (Dark Flow):} There are recent
indications that there is a large scale peculiar velocity flow with
amplitude larger than $400$km/sec on scales up to $100 h^{-1}$Mpc
($z\leq 0.03$) \cite{Watkins:2008hf} with direction $l \simeq
282^\circ \pm 11^\circ$, $b\simeq 6^\circ \pm 6^\circ$. Other
independent studies have also found large bulk velocity flows with
similar direction \cite{Lavaux:2008th} on scales of about $100
h^{-1}$Mpc or larger than $300 h^{-1}$Mpc
\cite{Kashlinsky:2008ut}. This large scale peculiar velocity flow is
known as {\it Dark Flow}.  The standard homogeneous-isotropic
cosmology (\lcdm), predicts significantly smaller amplitude and scale
of flows than what these observations indicate. The deviation of these
observations from $\Lambda$CDM  predictions is more than $3\sigma$.
Other studies \cite{Dai:2011xm} using Type Ia supernovae find a flow of a somewhat smaller magnitude, consistent with \lcdm . Even in these studies however the direction of the flow is similar to the direction found by Refs. \cite{Watkins:2008hf} and \cite{Kashlinsky:2008ut}. Thus, even though there is some controversy about the magnitude of the Dark Flow it appears that its direction is more robust even though its directional $1\sigma$ error region is probably larger than the one indicated in Ref. \cite{Watkins:2008hf} (for a more conservative $1\sigma$ error region see Ref. \cite{Dai:2011xm}).
 A possible connection of large scale velocity flows and cosmic
acceleration may be found in Ref. \cite{Tsagas:2009nh}.
\item
{\bf Fine Structure Constant Dipole:} Quasar absorber spectra
obtained using UVES (the Ultraviolet and Visual Echelle Spectrograph)
on the VLT (Very Large Telescope) in Chile and also previous
observations at the Keck Observatory in Hawaii \cite{King:2012id} have
indicated that the value of the fine structure constant at high
redshifts ($z\in [0.2,4.2]$) is not distributed isotropically. Its
anisotropy is well described by a dipole with axis directed towards $l
\simeq 320^\circ \pm 11^\circ$, $b\simeq -11^\circ \pm 7^\circ$. The
deviation of these observations from isotropy is
$4.1\sigma$~\cite{King:2012id,Mariano:2012wx}.
\item
{\bf Dark energy Dipole:} A recent fit of the Type Ia distance
moduli residuals (from the best fit \lcdm) to a dipole anisotropic
distribution has indicated~\cite{Mariano:2012wx} that the angular
distribution of these residuals is well described by a dipole
analogous to the Fine Structure Constant dipole. Its axis is towards
$l \simeq 309^\circ \pm 18^\circ$, $b\simeq -15^\circ \pm 11^\circ$
and deviates by only about $11^\circ$ from the fine structure constant
dipole. The deviation of these observations from isotropy is at the
$2\sigma$ level.
\end{enumerate}
Each one of the above observed deviations from isotropy is between
$2\sigma$ and $4\sigma$. The angular proximity of the corresponding
anisotropy axes makes their combination even more unlikely in an
isotropic universe where there is no correlation between them. In
Ref.~\cite{Mariano:2012wx} it was shown that the combined magnitude+alignment of the
fine structure constant $\alpha$ and dark energy dipoles has a probability less than one
part in $10^6$ to occur in an isotropic universe where the two dipoles
are uncorrelated.

A physical model was proposed in
Refs.~\cite{Mariano:2012wx,BuenoSanchez:2011wr} that has the potential
to explain the existence and the alignment of the above three
dipoles. The model is based on the existence of a topological defect
(e.\@g.\@ a global monopole) with Hubble scale core formed during a recent phase transition by an
$O(3)$ symmetric scalar field {\it non-minimally coupled to
electromagnetism}. An off-center observer with respect to the monopole
center would observe faster accelerating expansion towards the core
where the vacuum energy density is larger and also varying $\alpha$
along the same direction due to the variation of the scalar field
magnitude. This model is a generalization of `Topological Inflation'
\cite{topinfl} and has been called {\it Extended Topological
Quintessence}~\cite{Mariano:2012wx} due to its non-minimal coupling to
electromagnetism. The model also has some similarities with texture models which have been considered as a physical origin of the observed Cold Spots on CMB maps \cite{Cruz:2007pe}. In contrast to Extended Topological Quintessence however, texture models have not been considered as physical origin for cosmic accelerated expansion of for spatial variation of $\alpha$.

Extended Topological Quintessence makes the following qualitative
predictions~\cite{Mariano:2012wx}:
\begin{enumerate}
\item
{\bf Large Scale Velocity Flows:} Due to the stronger repulsive
gravity in the defect core a large scale peculiar velocity flow is
predicted along the axis that connects the off center observer and the
monopole core. The direction of the flow is predicted to be away from
the repelling core (`Great Repulser') and its scale is predicted to be
the Hubble scale (the defect core scale). A reversal of the velocity
flow direction is predicted for observations that go beyond the defect
core. As discussed in the following section, the direction of the
observed Dark Flow is consistent with the directions of the fine
structure and dark energy dipoles in accordance with the above
prediction.
\item
{\bf Correlation between values of $\alpha$ and presence of strong
magnetic fields:} As discussed in Ref.~\cite{Mariano:2012wx}, the
scalar field magnitude is expected to depend weakly of the presence of
local strong magnetic fields. This magnitude variation is in turn
expected to lead to local variations of $\alpha$ in cosmological
regions with large magnetic fields.
\item
{\bf Maximal large scale CMB variation towards the defect core:} Due
to the recent formation of the global defect, the ISW effect is
expected to lead to large temperature differences between opposite
directions in the sky along the direction towards the defect core. A
large part of this temperature asymmetry would have been subtracted
from CMB maps along with the dipole moment which is mainly due to our
motion with respect to the CMB. However, smaller traces of this
asymmetry could have survived the dipole subtraction and may be
detectable in large scale CMB maps.
\end{enumerate}

A wide range of {\it large scale anomalies} have been detected on CMB
maps~\cite{Copi:2010na,Bennett:2010jb}. The anomalies include an
abnormal alignment and planarity of the octopole and quadrupole
moments\cite{Tegmark:2003ve,Schwarz:2004gk}, the existence of two large and deep cold spots\cite{Bennett:2003bz,Cruz:2006fy,Mukherjee:2004in}, the lack of
large scale power\cite{Spergel:2003cb,Copi:2008hw,Ayaita:2009xm,Hajian:2007pi}, the even excess of the CMB power spectrum\cite{Kim:2010st}, the
hemispherical power asymmetry\cite{Hoftuft:2009rq} and quadrupolar dependence of the two
point function (see Ref.~\cite{Bennett:2010jb} for a detailed
review). Recent evidence for mirror symmetry and antisymmetry (along
different directions) has also been obtained~\cite{Finelli:2011zs,Land:2005jq}
using the ILC WMAP7 CMB map\cite{Jarosik:2010iu}. Finally evidence for the existence of statistically significant giant rings in the CMB sky has also been reported\cite{Kovetz:2010kv}.  Some of these anomalies appear to be related
to a large scale bipolar asymmetry of the CMB even though there is no
current quantitative study of a physical model than can give rise to
all these anomalies simultaneously (see however \cite{Ackerman:2007nb,Cruz:2007pe,Donoghue:2007ze,Erickcek:2008sm}). Due to the lack of such a model
these anomalies are usually assumed to be a posteriori manifestations
of expected large statistical fluctuations.

Having at our disposal a well defined physical model which makes specific
predictions allows us to focus on specific aspects of CMB maps and
search for signatures of our model. Thus, in what follows we focus on
the predicted large scale CMB anisotropy and search for the axis of
maximal temperature asymmetry in the WMAP7 ILC map. In particular we
consider three large scale Healpix pixelizations\cite{Gorski:2004by} of the WMAP7 ILC map
and identify those pairs of opposite pixels in the sky that correspond
to Maximum Temperature Difference. We compare the magnitude of this
Maximum Temperature Asymmetry (MTA) with that expected from an isotropic
model using Gaussian simulated CMB maps. We also compare the direction
of the MTA with the direction of the other
observed cosmic asymmetry axes (Dark Flow, Dark Energy Dipole and
$\alpha$ Dipole). We find the likelihood that the observed magnitude
and alignment would occur by chance in an isotropic model with no
correlation between the CMB and the other observables.

The structure of this paper is the following: In the next section we
describe in some detail the method for identifying the MTA magnitude and direction in the WMAP7 ILC map. We
also show the resulting magnitude and direction and also its alignment with
the other observed axes. We then compare the observed magnitude and
alignment with those obtained by $10^4$ Gaussian simulated ILC maps
based on \lcdm. In section III we discuss the implications of our
results and point out the next steps of this research program.
\begin{figure*}[t]
\centering
\includegraphics[scale=.5]{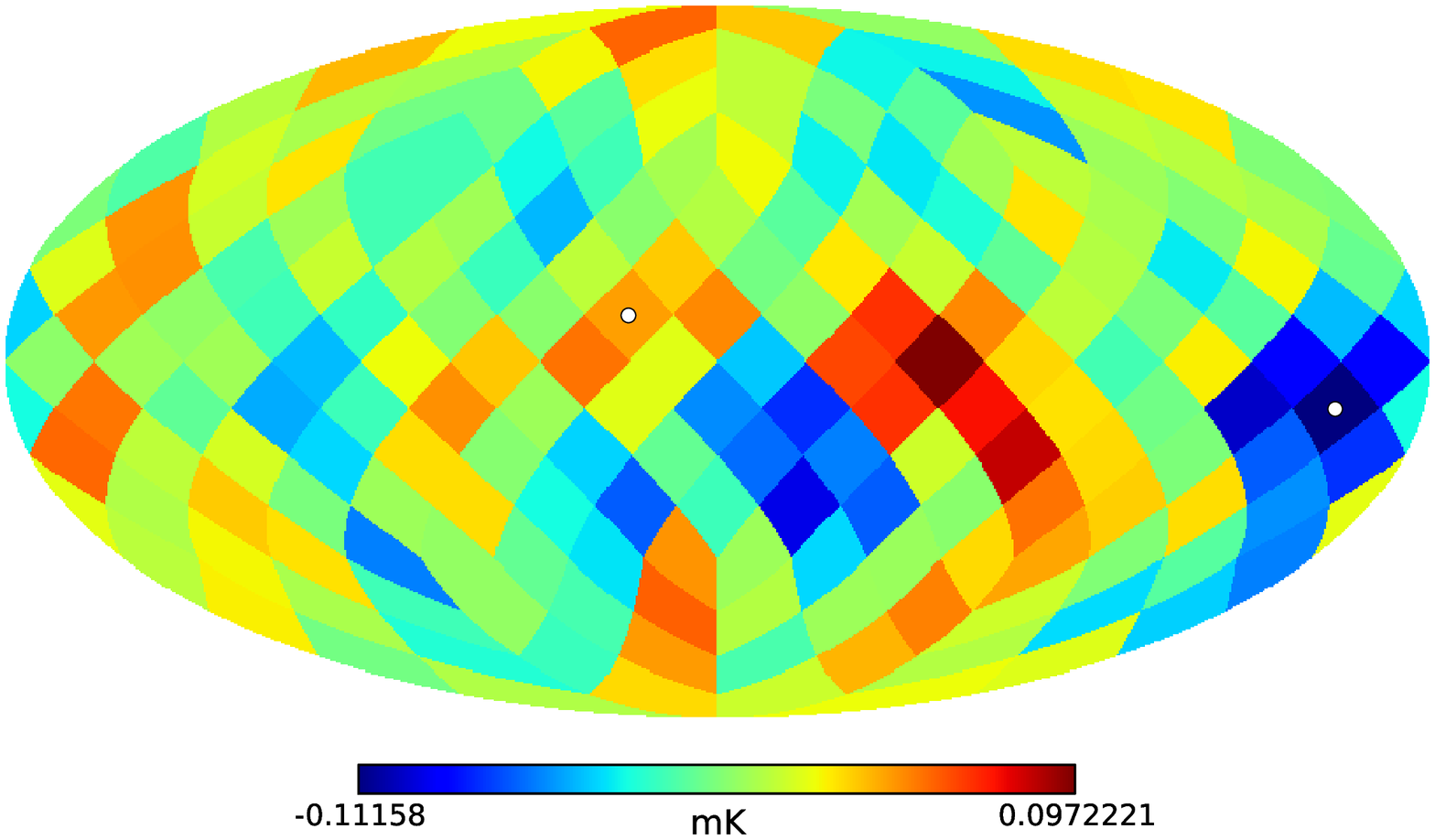}\vspace{.4cm}\\
\includegraphics[scale=.5]{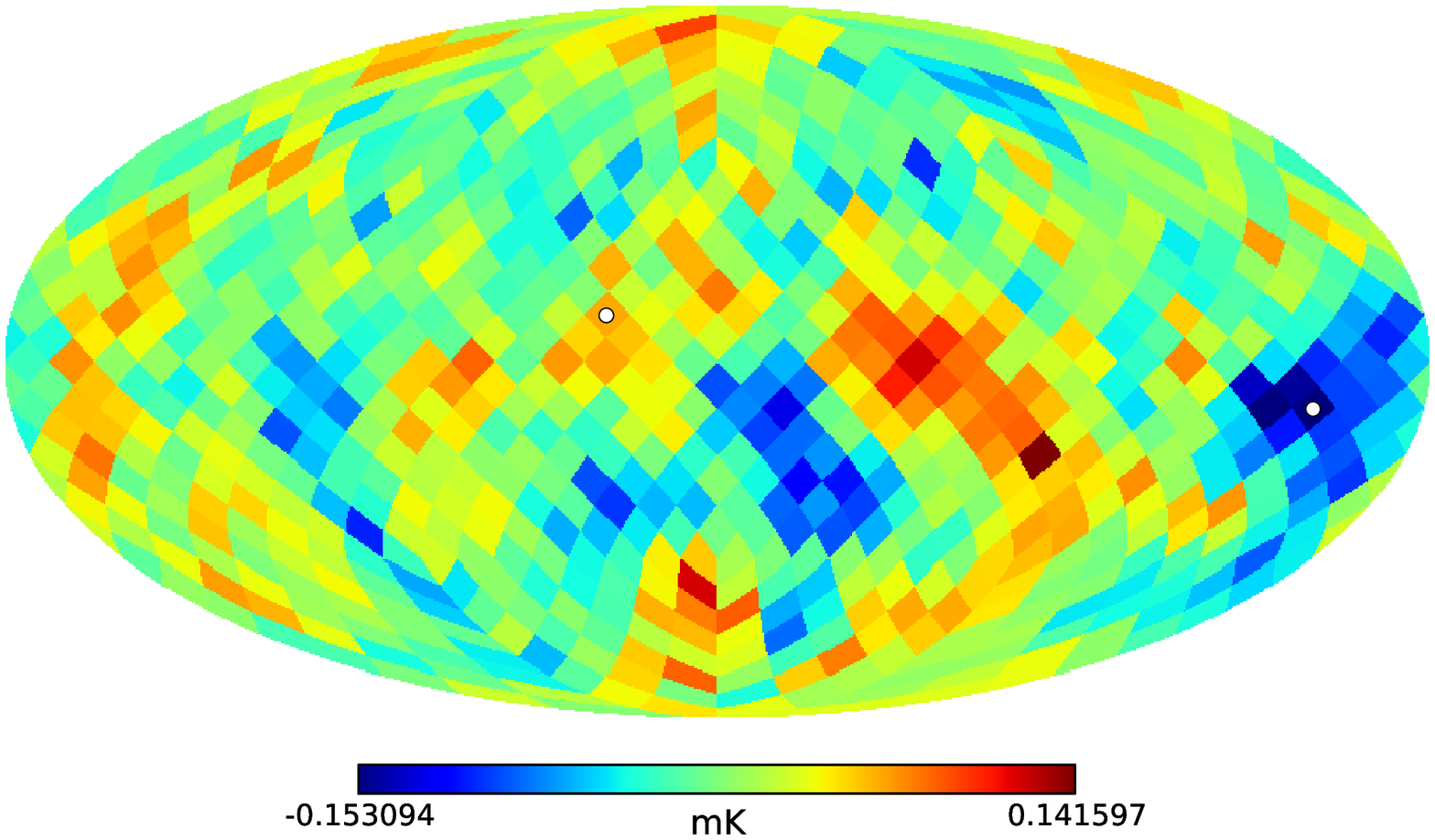}\vspace{.4cm}\\
\includegraphics[scale=.5]{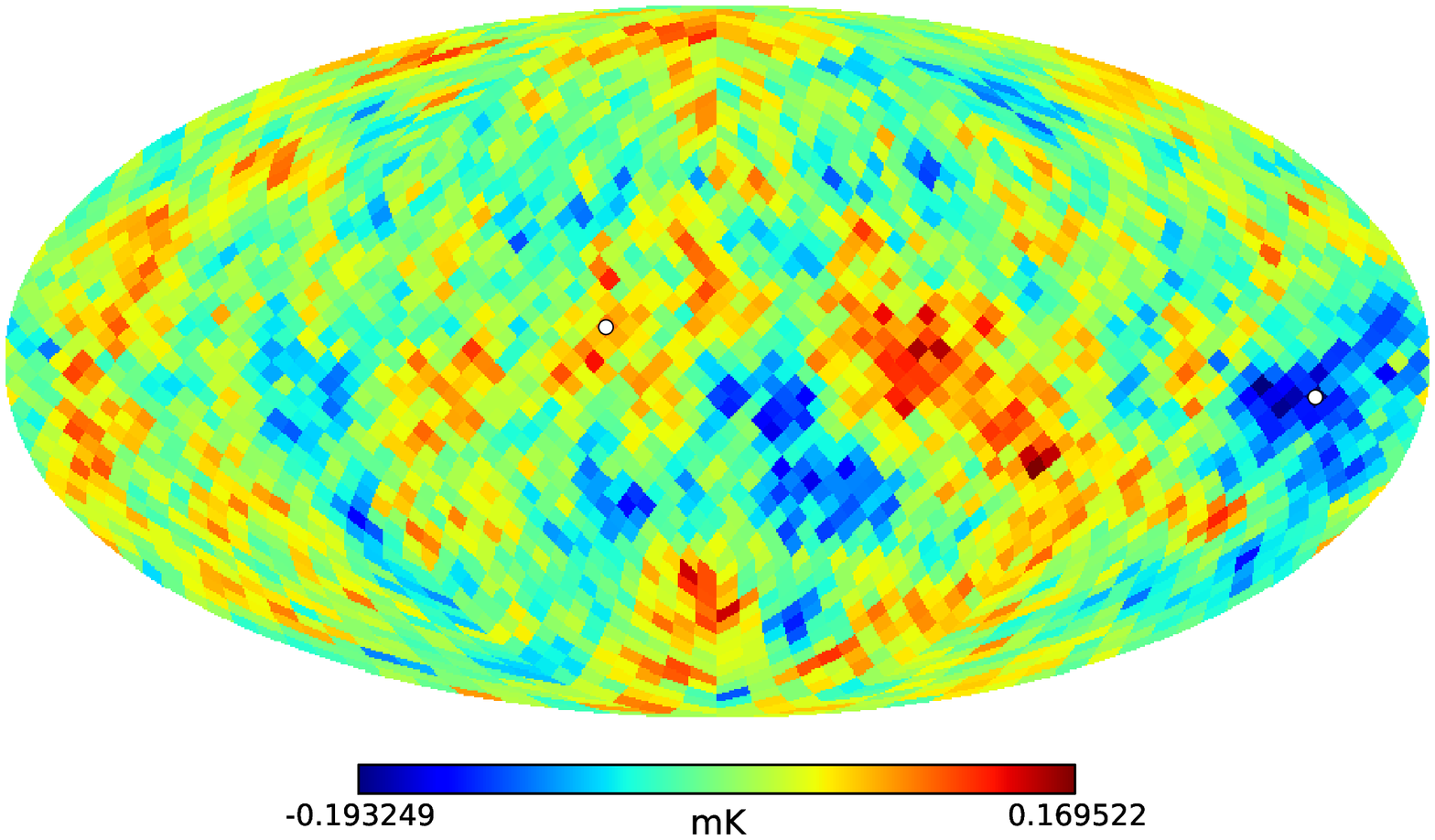}
\caption{Degraded temperature maps obtained from the 7 years ILC CMB
map with $N_{side}=4,8,16$. The white dots define the maximum
temperature differences direction. }
\label{fig:dag_maps}
\end{figure*}

\section{Method-Results}
The subtraction of the dipole moment from the CMB maps removes along
with the dominant Doppler component any cosmological signal that may
happen to have a dipole anisotropy. Such a signal is expected to
emerge in the context of extended topological quintessence as
discussed in the Introduction (see also
Refs.~\cite{Mariano:2012wx,BuenoSanchez:2011wr}). In addition to the
dipole however, an off center observer will also experience axial
anisotropies corresponding to higher moments although at a smaller
magnitude \cite{Grande:2011hm,Alnes:2006pf}. Depending on the dynamics
and the geometry of the forming defect these higher moment asymmetries
may have detectable magnitude. Such asymmetry could manifest itself as
maximized temperature difference between opposite pixels in the CMB
sky.  In order to obtain the direction and magnitude of such residual
MTA we use the following steps applied on
the WMAP7 foreground reduced ILC map pixelized according to
Healpix. In order to minimize foreground contamination we focus on
large angular scales ( $N_{side}=4$ (pixel size about $14.7^\circ$),
$N_{side}=8$ (pixel size about $7.3^\circ$), $N_{side}=16$ (pixel size
about $3.7^\circ$))

\begin{enumerate}
\item
Construct a Temperature Difference Map (TDM) obtained by assigning to
each pixel a number equal to the difference between its temperature
value and the value of the temperature of the opposite pixel in the
sky. Thus we have
\be
D^{-} (\hat n_i) =
\left( {T} (\hat n_i) - { T}(-\hat n_i) \right)
\label{deftdm}
\, ,
\ee
where $\hat n_i$ is the direction of the $i^{th}$ Healpix pixel. A similar estimator was considered in Ref. \cite{Naselsky:2011ec} in an effort to test statistical isotropy of CMB maps. In
the context of the Healpix pixelization the opposite pixel is always
simply defined and identified. By construction, opposite pixels of the TDM are
assigned to opposite values.
\item
In the TDM we select the pixel $D^{-}_{max} (\hat n_k)$ with the maximum
absolute value. This pixel along with the pixel located opposite to it
defines the axis of MTA. If the dipole
had not been subtracted the MTA axis would be almost identical to the dipole
axis. Thus the MTA axis is the residual asymmetry axis after the
subtraction of the dipole.
\item
The direction of the MTA is then compared with the directions of other
cosmic asymmetry axes ($\alpha$ dipole, Dark Energy dipole and Dark
Flow) and the corresponding angular differences are identified.
\item
The magnitude and direction of the MTA are compared with a large
number of $\Lambda$CDM simulated ILC maps \cite{Eriksen:2005kg} and we
evaluate the likelihood to obtain the observed MTA magnitude (or
larger) in the context of \lcdm.  The likelihood of obtaining the
angular differences (or smaller) with the other cosmic asymmetry axes in the
context of $\Lambda$CDM is also evaluated.
\end{enumerate}

The WMAP7 ILC maps using Healpix
pixelizations with $N_{side}=4,8,16$ (corresponding to a pixel size of
$\sqrt{4\pi/\left(12N_{side}^2\right)}$ rad i.\@e.\@ about
$14.7^\circ$, $7.3^\circ$ and $3.7^\circ$ respectively) are shown in
Figure~\ref{fig:dag_maps} along with the MTA pixels. The original map
has $N_{side}=512$.  The proximity of the MTA axis with one of the Cold Spots center
is evident.

\begin{figure*}[t]
\centering
\includegraphics[scale=1]{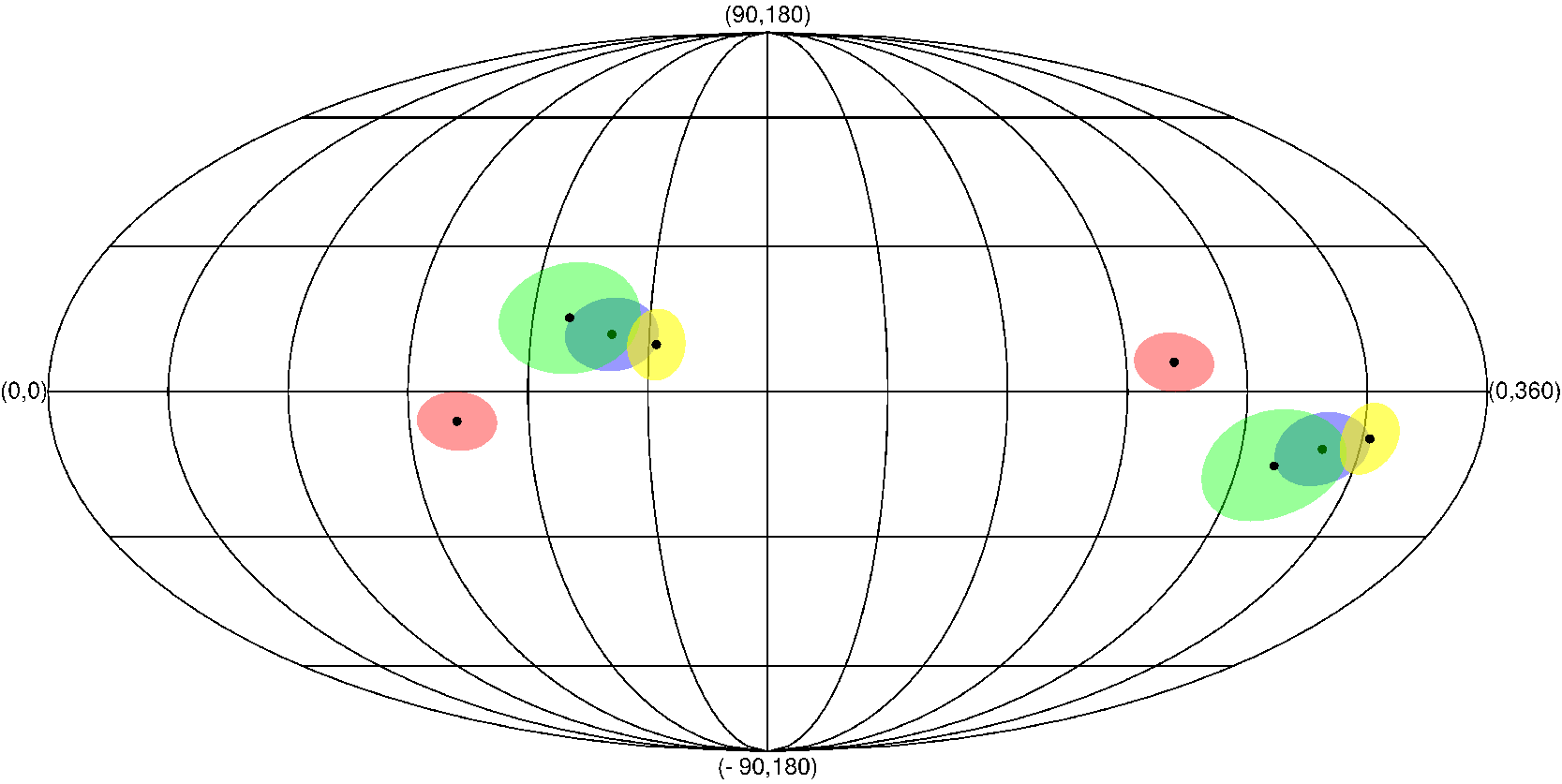}
\caption{Directions in galactic coordinates for the $\alpha$ (blue)
  and Dark Energy (green) dipoles, for the Dark Flow direction (red)
  and for the direction of MTA in the 7
  years ILC CMB map degraded to $N_{side}=8$ (yellow). The opposite corresponding directions are also shown.}
\label{fig:error_blobs}
\end{figure*}

In Table~\ref{tab:directions} we show the directions in galactic
coordinates of the four cosmic asymmetry axes.
\begin{table}[t]
\centering
\begin{tabular}{|l|c|c|}
\hline
                    & $l$ ($^\circ$)   & $b$ ($^\circ$)   \\
\hline
MTA ($N_{side}=4$)  & $337.5 \pm 14.7$  & $-9.6 \pm 14.7$   \\
MTA ($N_{side}=8$)  & $331.9 \pm 7.3$  & $-9.6 \pm 7.3$   \\
MTA ($N_{side}=16$) & $331.9 \pm 3.7$  & $-7.2 \pm 3.7$   \\
$\alpha$ dipole     & $320.5 \pm 11.8$ & $-11.7 \pm 7.5$  \\
Dark Energy dipole  & $309.4 \pm 18.0$ & $-15.1 \pm 11.5$ \\
Dark Flow direction & $282 \pm 11$     & $6 \pm 6$        \\
\hline
\end{tabular}
\caption{Directions in galactic coordinates for the $\alpha$ \cite{King:2012id,Mariano:2012wx}and Dark
  Energy dipoles\cite{Mariano:2012wx}, the Dark Flow and the maximum CMB temperature
  difference (MTA). For the Dark Flow direction we have used Ref. \cite{Watkins:2008hf}. The larger scale direction of Ref. \cite{Kashlinsky:2008ut} is consistent with that of Ref. \cite{Watkins:2008hf} but it has significantly larger errorbars. The error on the MTA direction has been taken to
  be equal to the side of the pixel,
  $\sqrt{4\pi/\left(12N_{side}^2\right)}$ rad.}
\label{tab:directions}
\end{table}
In Figure~\ref{fig:error_blobs} we show these directions in a Mollweide
projection. The filled contours around each direction correspond to
the $1\sigma$ error regions.
In Table \ref{tab:angular_dists} we show the corresponding angular
separations for each pair.
\begin{table}[t]
\centering
\begin{tabular}{|l|c|c|c|c|}
\hline
                & MTA      & $\alpha$ dipole & DE dipole     & DF direction  \\
\hline
MTA  ($N_{side}=8$)& 0.0        & $11.4 \pm 12$   & $22.6 \pm 18$ & $52.1 \pm 11$ \\
$\alpha$ dipole & $11.4 \pm 12$ & 0.0             & $11.3 \pm 18$ & $42.2 \pm 11$ \\
DE dipole       & $22.6 \pm 18$ & $11.3 \pm 18$       & 0.0            & $34.4 \pm 18$ \\
DF direction    & $52.1 \pm 11$ & $42.2 \pm 11$        & $34.4 \pm 18$      & 0.0             \\
\hline
\end{tabular}
\caption{Angular distances in degrees between the Alpha and Dark Energy dipoles,
  the Dark Flow and the MTA
  directions. For the MTA direction we have chosen the result
  obtained in the $N_{side}=8$ case.}
\label{tab:angular_dists}
\end{table}

\begin{figure*}[t]
\centering
\includegraphics{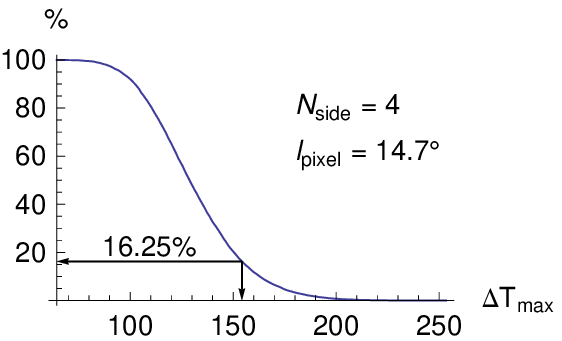}
\includegraphics{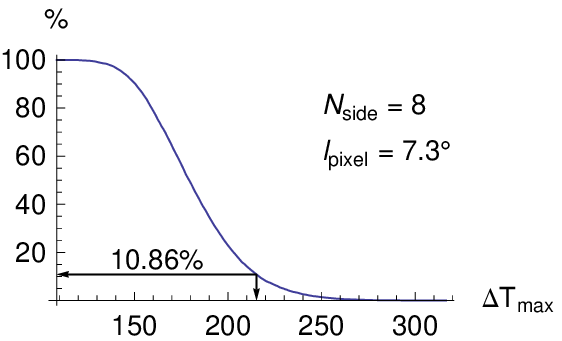}
\includegraphics{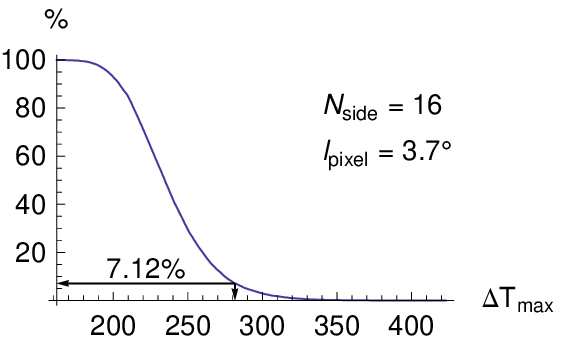}
\caption{Percentage of the maximum temperature difference values
  obtained from the simulated maps bigger than the observed maximum
  temperature difference obtained from the degraded maps with
  $N_{side}=4,8,16$.}
\label{fig:diffs_prob}
\end{figure*}

The cumulative probability for obtaining a given value of MTA or
larger, may be obtained using $10^4$ simulated statistically isotropic
\lcdm~ILC maps \cite{Eriksen:2005kg}. The result is shown in Figure
\ref{fig:diffs_prob} for each one of the three angular resolutions
(pixel sizes) considered.
We used the publicly available
\lcdm~simulated ILC maps of Ref.~\cite{Eriksen:2005kg}. The observed
value of the MTA magnitude is indicated by an arrow. The probability
to obtain the observed magnitude of MTA (or larger) in the context of
$\Lambda$CDM varies between $16\%$ and $7\%$ depending on the ILC map angular
resolution. This result by itself does not indicate any statistically
significant deviation from \lcdm~predictions. Perhaps, this is the
main reason that this simple statistic has been largely ignored by
previous studies (see however Ref. \cite{Naselsky:2011ec}). However, the statistic becomes more interesting when
the proximity of the direction of the MTA to other cosmic asymmetry
axes is considered.

\begin{figure*}[t]
\centering
\includegraphics{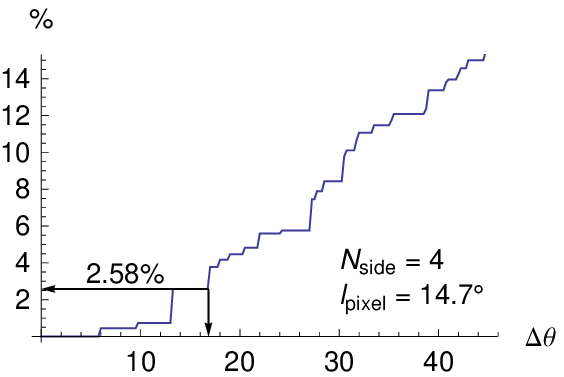}
\includegraphics{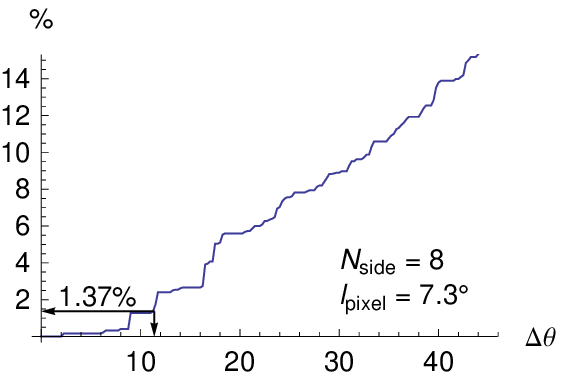}
\includegraphics{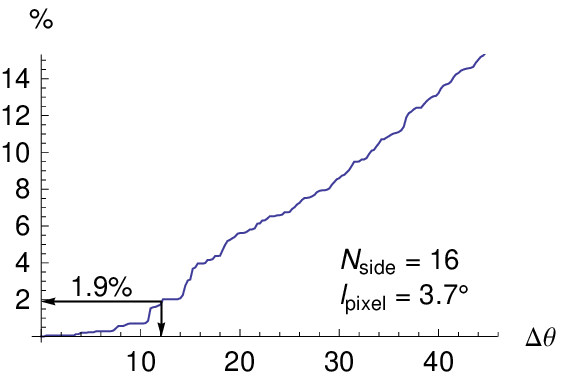}
\caption{Percentage of the maximum temperature difference directions
  obtained from the simulated maps closer to the $\alpha$ dipole than
  the observed maximum MTA obtained from the
  degraded maps with $N_{side}=4,8,16$.}
\label{fig:angdists_alpha_prob}
\end{figure*}

\begin{figure*}[t]
\centering
\includegraphics{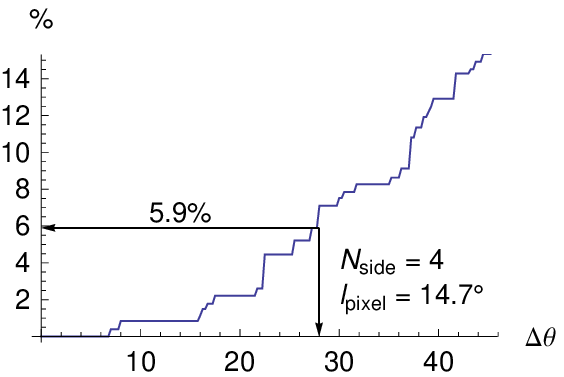}
\includegraphics{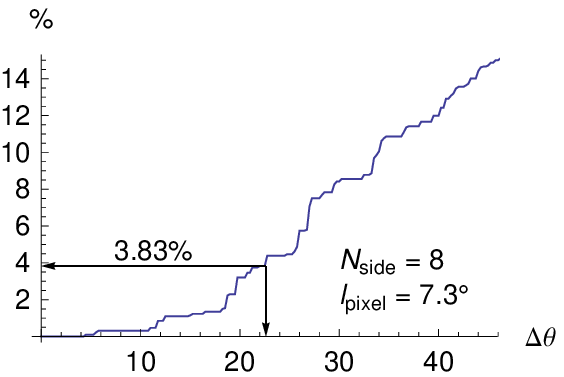}
\includegraphics{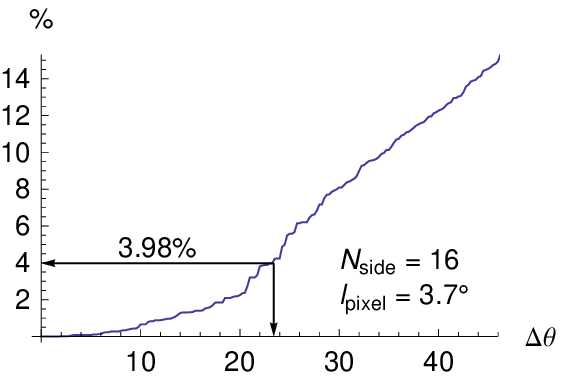}
\caption{Percentage of the MTA directions
  obtained from the simulated maps closer to the Dark Energy dipole than
  the observed MTA directions obtained from the
  degraded maps with $N_{side}=4,8,16$.}
\label{fig:angdists_DE_prob}
\end{figure*}

\begin{figure*}[t]
\centering
\includegraphics{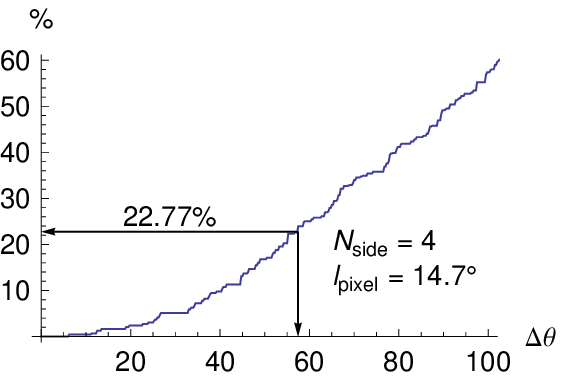}
\includegraphics{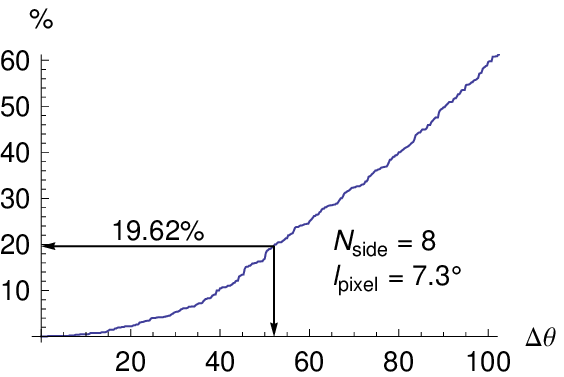}
\includegraphics{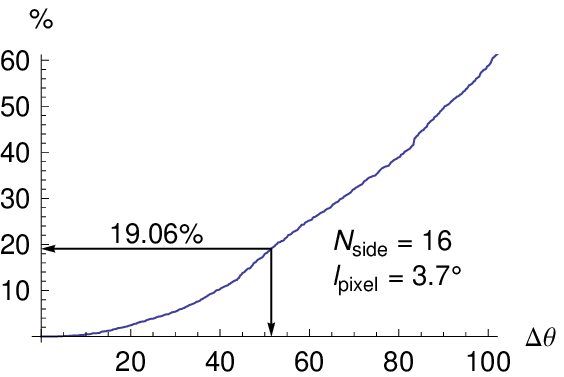}
\caption{Percentage of the MTA directions
  obtained from the simulated maps closer to the Dark Flow direction than
  the observed MTA directions obtained from the
  degraded maps with $N_{side}=4,8,16$.}
\label{fig:angdists_DF_prob}
\end{figure*}

In Fig.~\ref{fig:angdists_alpha_prob}, we plot the percentage of the
MTA directions obtained from the simulated maps that form an angle
with the observed $\alpha$ dipole direction smaller than a given angle
(shown on the horizontal axis). The cases of map resolutions
corresponding to $N_{side}=4,8,16$ are shown. The angle between the
observed MTA and the {\it observed} $\alpha$ dipole direction is indicated
by an arrow on each plot.
In Figs.~\ref{fig:angdists_DE_prob}, \ref{fig:angdists_DF_prob} we
show the corresponding plots where instead of the $\alpha$ dipole
direction we have used the Dark Energy dipole and Dark Flow directions
respectively.


The probability that the \lcdm~simulated maps reproduce the observed
alignment of cosmic asymmetries varies between $1.37\%$ (alignment
with $\alpha$ dipole), and $22.77\%$ (alignment with Dark Flow). The
combined probability to obtain both a large enough magnitude and
angular proximity of MTA to the $\alpha$ dipole direction is shown in
Table~\ref{tab:mag_alpha_probs}.
\begin{table*}[t]
\centering
\begin{tabular}{|l|c|c|c|}
\hline
                                                       & $MTA_{sims} > MTA_{obs}(\%)$ &
$\theta_{MTA-\alpha,sims}<\theta_{MTA-\alpha,obs}(\%)$ &
both$(\%)$                                                                                   \\
\hline
$N_{side}=4$                                           & 16.25
                                                       & 2.58                         & 0.48 \\
$N_{side}=8$                                           & 10.86
                                                       & 1.37                         & 0.19 \\
$N_{side}=16$                                          & 7.12
                                                       & 1.9                          & 0.12 \\
\hline
\end{tabular}
\caption{Probabilities of obtaining a simulated CMB map with a maximum
temperature difference bigger than the observed one and with a MTA
direction closer to the $\alpha$ dipole direction than the observed
one.}
\label{tab:mag_alpha_probs}
\end{table*}
In particular, the probability to
obtain the observed MTA magnitude (or larger) and the observed angular
proximity to the $\alpha$ dipole direction in the context of
\lcdm~varies between $0.5\%$ and $0.1\%$ depending on the angular
resolution of the WMAP7 ILC map. If the probability of obtaining the
$\alpha$ dipole magnitude in the context of \lcdm~is also factored in,
the probability reduces to about one part in $10^7$ which is similar
to the probability for obtaining simultaneously the Dark Energy and
the $\alpha$ dipoles in the observed directions~\cite{Mariano:2012wx}.

The last column of Table~\ref{tab:mag_alpha_probs} is also shown in
Table~\ref{tab:mag_dips_probs} along with the corresponding results
for the other two anisotropy directions corresponding to the Dark
energy Dipole and the Dark Flow.

\begin{table}[t]
\centering
\begin{tabular}{|l|c|c|c|}
\hline
              & $\alpha(\%)$ & DE$(\%)$ & DF$(\%)$ \\
\hline
$N_{side}=4$  & 0.48         & 0.95     & 3.37     \\
$N_{side}=8$  & 0.19         & 0.52     & 2.06     \\
$N_{side}=16$ & 0.12         & 0.28     & 1.28     \\
\hline
\end{tabular}
\caption{Probabilities of obtaining a simulated CMB map with both a maximum
  temperature difference bigger than the observed one and a
  MTA direction closer to the $\alpha$,
  Dark Energy dipole and Dark Flow directions than the observed one.}
\label{tab:mag_dips_probs}
\end{table}

We stress that the above abnormally low probabilities assume that the
corresponding datasets (Keck+VLT quasar absorbers \cite{King:2012id},
Dark Flow data \cite{Watkins:2008hf,Kashlinsky:2008ut}, Union2 data
\cite{Amanullah:2010vv} and ILC maps \cite{Jarosik:2010iu}) are free
of systematic errors. The potential validity of these datasets combined with the generic nature of the statistical tests
applied, assigns a particularly low likelihood to the statistical
isotropy feature of \lcdm.

Nevertheless, the existence of a physical
model where the alignment of the above axes will appear with a
significantly larger probability is a prerequisite before putting
\lcdm~to disfavor. Even though the qualitative predictions of Extended
Topological Quintessence appear to be significantly more consistent
with the observed cosmic asymmetries than \lcdm, a quantitative
analysis is required before any valid conclusion in favor of Extended
Topological Quintessence is drawn. Such an analysis is currently in
progress.

\section{Conclusion-Outlook}
We have identified a direction on Maximum Temperature Asymmetry (MTA)
of the WMAP7 foreground reduced ILC map. Even though the magnitude of
this asymmetry is consistent with \lcdm~at the $2\sigma$ level, its
direction is abnormally close to other observed cosmic asymmetry
axes. The direction of the MTA is close to the direction of
one of the Cold Spots. This angular proximity may imply that this Cold
Spot (or the opposite located Hot Region) is physically related to the
existence of other cosmic asymmetry axes. In the context of Extended
Topological Quintessence, the existence of such a feature (Hot or Cold
spot) is expected to exist at the core of the `Great Repulser' global
defect while in the opposite direction an opposite temperature behavior is expected.

The planarity and alignment of the CMB octpulole and quadrulole moments may be partly due to a combination of two or more features on the preferred plane of these moments. Indeed, MTA axis we have identified lies on this preferred plane and therefore the MTA may be related to the observed quadrupole-octupole alignment \cite{Tegmark:2003ve}

An interesting extension of this project is the derivation of the
detailed CMB signature predicted by Extended Topological
Quintessence. Such a derivation would involve a cosmological
simulation of the evolution of the non-minimally coupled $O(3)$ scalar
field that gives rise to the recent formation global defect. The
linear metric perturbations that emerge due to this formation can then
be numerically calculated~\cite{Bennett:1992fy} and the corresponding
ISW effect can be derived in a straightforward manner. This numerical
analysis can also lead to detailed predictions about the magnitude and
geometry of the other cosmic asymmetry axes (Dark Flow, Dark Energy
and $\alpha$ dipole). This analysis is currently in progress.
\newpage

{\bf Numerical Analysis Files:} The data, Mathematica and Healpix program
files used for the numerical analysis files may be downloaded from
\url{http://leandros.physics.uoi.gr/mta}.

\section*{Acknowledgments}
We acknowledge the use of the Legacy Archive
for Microwave Background Data Analysis (LAMBDA)\cite{lambda}
 and the use of the HEALPix package\cite{Gorski:2004by}.
This research has been co-financed by the European Union (European
Social Fund - ESF) and Greek national funds through the Operational
Program ``Education and Lifelong Learning'' of the National Strategic
Reference Framework (NSRF) - Research Funding Program: ARISTEIA.\@
Investing in the society of knowledge through the European Social
Fund.

\end{document}